\documentclass[conference]{IEEEtran}
\IEEEoverridecommandlockouts
\usepackage{amssymb}
\usepackage{amsmath}      
\usepackage{cite}         
\usepackage{algorithmicx} 
\usepackage{graphicx}     
\usepackage{tabularx, booktabs} 
\usepackage{nomencl}
\makenomenclature
\usepackage{subfig}
\usepackage{caption}
\usepackage{xcolor} 
\usepackage{color}
\usepackage{hyperref}
\usepackage{multirow}
\DeclareSubrefFormat{parens}{#1(#2)}
\usepackage{tablefootnote}

\begin{document}

\title{Configuration and EMT Simulation of the 240-bus MiniWECC System Integrating Offshore Wind Farms (OWFs)\\
\thanks{This work is funded by the Laboratory Directed Research and Development
(LDRD) at the Pacific Northwest National Laboratory (PNNL) as part of the E-COMP initiative. PNNL is operated by Battelle for the DOE under Contract DOE-AC05-76RL01830}
}

\author{
\IEEEauthorblockN{Buxin She, Hisham Mahmood, Marcelo Elizondo, Veronica Adetola
}
\IEEEauthorblockA{\textit{Parcific Northwest National Laboratory} \\
Richland, USA\\
\{buxin.she, hisham.mahmood, marcelo.elizondo, veronica.adetola\}@pnnl.gov}
\and
\IEEEauthorblockN{Yuqing Dong}
\IEEEauthorblockA{\textit{The University of Tennessee Knoxville}\\
Knoxville, USA \\
ydong22@utk.edu}
}

\maketitle

\begin{abstract}
As offshore wind farms (OWFs) become increasingly prevalent in Northern California and Southern Oregon, they introduce faster dynamics into the Western Electricity Coordinating Council (WECC) system, reshaping its dynamic behavior. Accordingly, electromagnetic transient (EMT) simulation is essential to assess high frequency dynamics of the WECC system with integrated OWFs. Against this background, this paper presents the integration of detailed dynamic models of OWFs into a 240-bus miniWECC system in PSCAD software. The sequential initialization technique is employed to facilitate the smooth initiation of a large-scale system in an EMT simulation. 
The performance of the configured model is assessed under wind speed variations and grounded faults, demonstrating the effectiveness of the miniWECC system with OWFs. This system serves as a valuable basic use case for validating the fast dynamic performance of future WECC systems with high penetration of wind energy.
\end{abstract}

\begin{IEEEkeywords}
240-bus miniWECC system, EMT simulation, offshore wind
farms, inverter-based resources, PSCAD
\end{IEEEkeywords}

\section{Introduction}

For the benefit of Americans and the environment worldwide, the Biden's Administration has proposed the goal of achieving carbon pollution-free electricity by 2035. The broader objective is to ``deliver an equitable, clean energy future and put the United States on a path to achieve net-zero emissions, economy-wide, by no later than 2050." In response to this administration, the integration of offshore wind farms (OWFs) into power systems is experiencing significant expansion \cite{lesser2022biden}, particularly in regions like Northern California and Southern Oregon. 

The development of OWFs is crucial for the Western Electricity Coordinating Council (WECC) system. Pacific Northwest National Laboratory (PNNL) has conducted research on the OWFs development strategy to optimize the electrical system benefits \cite{travis2023}. The study identifies transmission line capacity as a key factor limiting the integration of OWFs on the West Coast.
A well-designed interregional transmission system utilizing high voltage alternating current (HVAC) \cite{chou2012comparative}, high voltage direct current (HVDC) \cite{sun2021cross}, and multi-terminal high voltage direct current (MTDC) \cite{liao2023stability} proves to be effective in addressing this limitation.
As a result, the net revenue could be maximized while also contributing to the development of carbon pollution-free electrical power grids.

As power electronic interfaced generation, OWFs significantly impact system dynamics, particularly when interacting with other inverter-based resources (IBRs) \cite{she2023inverter}. The occurrence of low-frequency oscillations at 3-4 Hz in ERCOT under weak grid conditions has been observed \cite{fan2018modeling}. Additionally, OWFs can be a source of high-frequency oscillations up to 2500 Hz \cite{zong2021grey}. These phenomena pose a threat to the secure operation of power systems integrated with OWFs. 
Therefore, beyond the economic analysis presented in \cite{travis2023}, it is essential to analyze the WECC system dynamics \cite{ali2021offshore}. Specifically, electromagnetic transient (EMT) simulation becomes a crucial tool for OWFs \cite{lin2018exact}, given its high accuracy and capability to capture intricate details of high-frequency dynamics.

Several studies have focused on the development and dynamic analysis of the WECC system. The early versions of the miniWECC system included a 179-bus model \cite{jung2002adaptation} and a 225-bus model \cite{yu2009evaluation}. 
Thereafter, targeting market design, a 240-bus miniWECC system with low penetration of renewables is developed in \cite{price2011reduced}. Following this, the 240-bus system is augmented \cite{yuan2020developing} with additional IBRs based on actual WECC data from 2018, configured in PSS/e with IBR dynamic models. It is later converted into a PSCAD case using PRSIM software and self-developed dynamic modules \cite{wang2022developing}. However, a synthetic system that integrates both the IBRs and OWFs has yet to be developed. 

The case developed in PSS/e \cite{yuan2020developing} is openly accessible but the system in PSCAD \cite{wang2022developing} is not available for public access as of the preparation of this paper. In this paper, the PSCAD model based on the work in \cite{yuan2020developing} is developed, and the OWFs are modeled in detail and integrated into the base case.
This work provides a foundational use case, essential for validating the fast dynamic performance of future WECC systems as they incorporate a growing share of offshore wind energy.

The rest of this paper is organized as follows: Section II introduces the basic miniWECC system, followed by key cascaded initialization techniques. Section III introduces the strategies for modeling, configuring, and integrating OWFs. Then, Section IV presents the time-domain simulation results, followed by conclusions drawn in Section V.

\section{Basic 240-bus MiniWECC System for EMT Simulation}

This section briefly introduces the basic 240-bus miniWECC system. It details the transformation process from the PSS/e model using E-Tran software \cite{cui2019electromechanical} and outlines the sequential initialization of the finalized PSCAD modules.

\subsection{MiniWECC System Overview}

\begin{figure*}[htbp]
 \centering
   \includegraphics[width=1\linewidth]{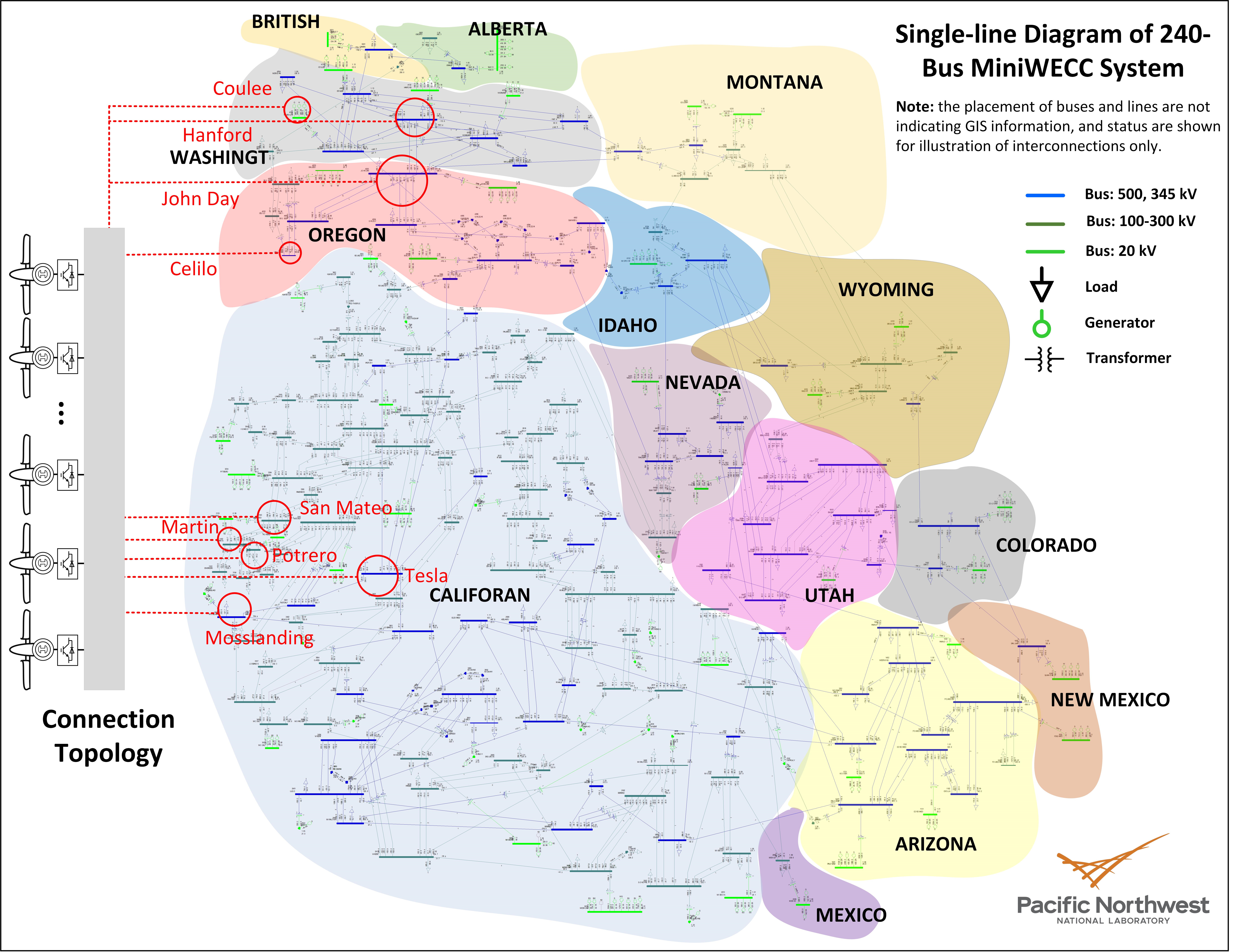}
 \caption{Diagram of 240-bus miniWECC system with integrated OWFs.}
 \label{fig:wecc}
\end{figure*}

Referring to the data in \cite{yuan2020developing}, the total generation capacity of the 240-bus miniWECC system is 291 gigawatts (GW), of which there are 59 GW grid-following (GFL) IBRs, including utility PV resources, wind, and distribution PV. More detailed data can be found in \cite{yuan2020developing, wang2022developing}. 
Fig. \ref{fig:wecc} shows the single-line diagram of the 240-bus miniWECC system, emphasizing the potential locations where the OWFs are installed and validated \cite{travis2023}.

\subsection{Configuration of the Basic PSCAD Case}

\subsubsection{SG-based miniWECC system}

\begin{figure}[htbp]
   \centering
   \subfloat[\label{fig:etran_component}]{
        \includegraphics[scale=0.75]{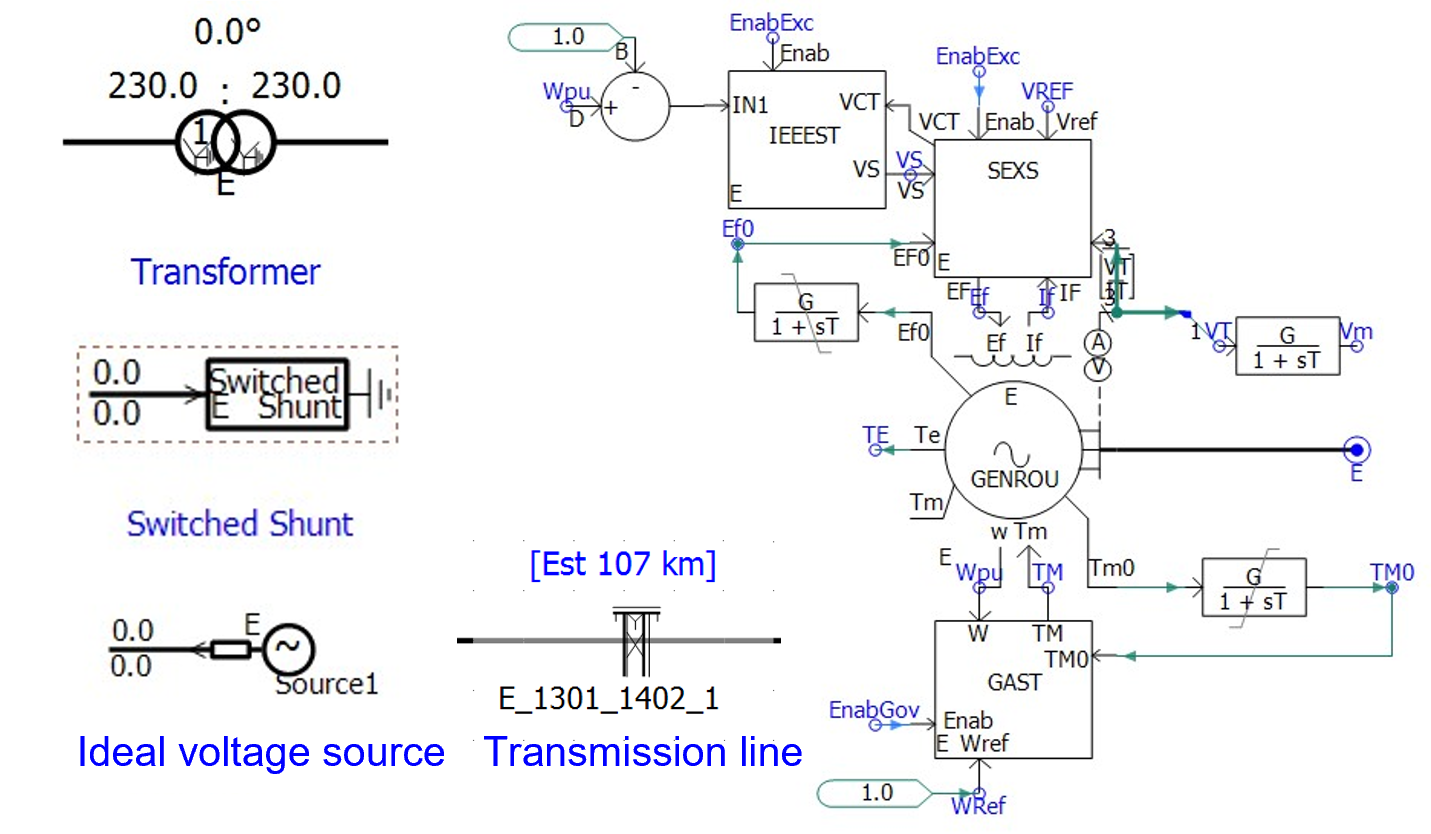}}
        \vspace{-0em} 
   \hfill
   \subfloat[\label{fig:etran_area}]{
         \includegraphics[scale=0.27]{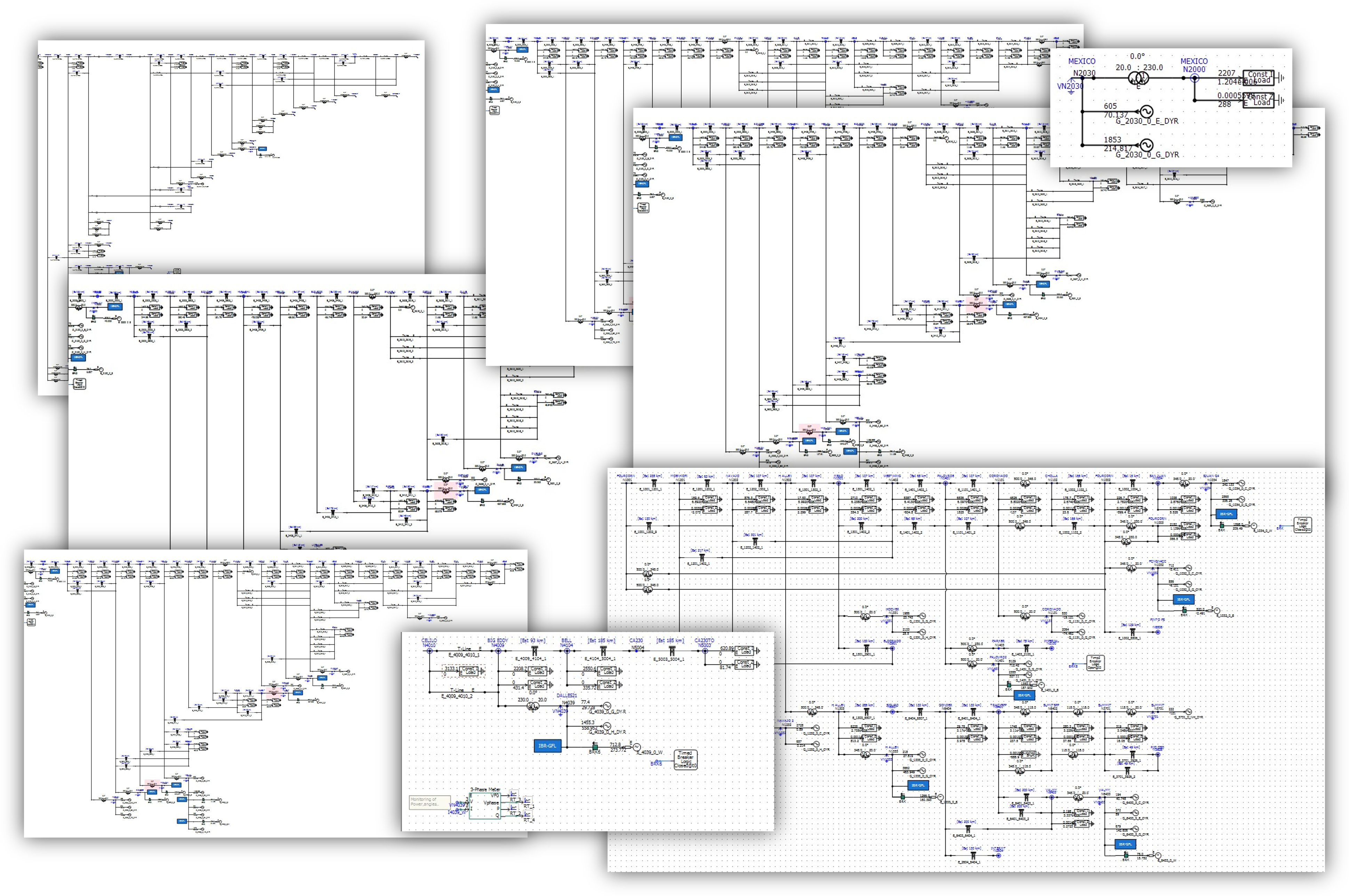}}
    \hfill
   \subfloat[\label{fig:GFL-IBR}]{
         \includegraphics[scale=0.48]{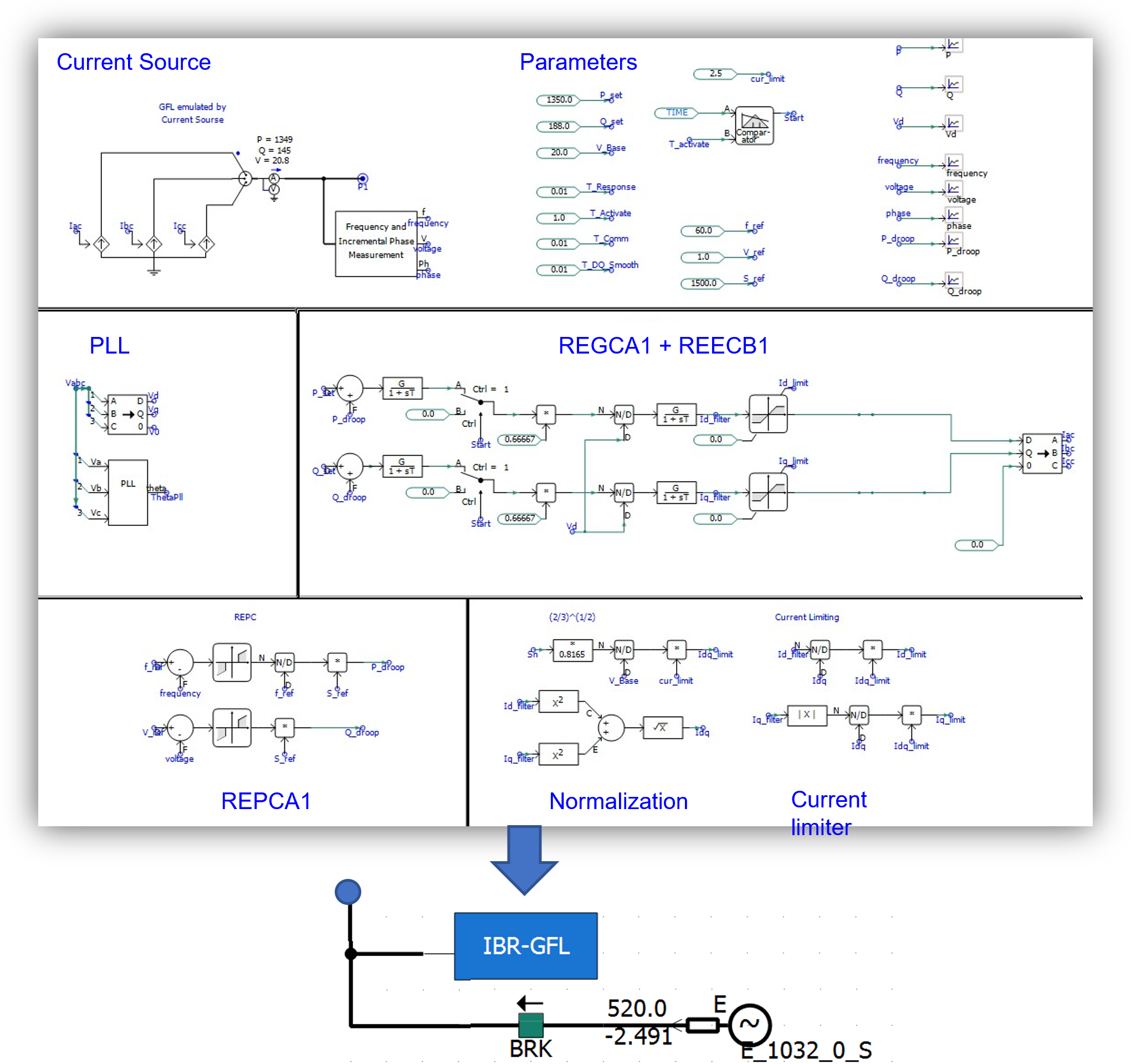}}
\caption{Diagram of PSCAD modules: (a) basic components converted from E-Tran; (b) interconnected areas of miniWECC system converted from E-Tran; (c) self-configured GFL IBR in parallel with ideal voltage source.}
\label{fig:etran}
\end{figure}

In this paper, E-Tran converts the basic PSS/e model into the PSCAD format, thereby avoiding the process of manually deploying transmission lines, loads, and synchronous generators (SGs). E-Tran features built-in dynamic libraries of exciters and governors for SGs. Fig. \ref{fig:etran_component} shows some PSCAD components in the E-Tran library. 

In comparison to PRSIM, E-Tran is more effective in converting PSS/e cases, generating more organized single-line diagrams which expedite the development process. 
Fig. \ref{fig:etran_area} visualizes the partitioned 8 areas within the 240-bus miniWECC system in PSCAD. 
It's important to note that E-Tran lacks an IBR library, and each IBR is represented by the ideal voltage source. Therefore, the miniWECC system obtained from E-Tran remains an SG-based system. The next step involves modeling IBRs and integrating them into the transformed case.

\subsubsection{Configuration of grid-following inverters}

Following the practice in \cite{wang2022developing}, PSCAD modules are manually constructed for grid-following (GFL) IBRs featuring dynamic models of REGCA1, REECB1, and REPCA1 shown in Fig. \ref{fig:GFL-IBR}. The REGCA1 model is modified wherein the \textit{d}- and \textit{q}-axis currents are transformed into a three-phase current using the phase lock loop. In addition, the frequency droop and voltage droop deadbands are established at $\pm 0.017$ Hz and $\pm 0.01$ p.u., respectively. Then, the next stage is to integrate GFL IBRs into the SG-based miniWECC system.

\subsection{Sequential Initilization}

Initiating EMT simulation from zero states, particularly for a large-scale system, poses considerable challenges. Therefore, the sequential initialization technique is employed to ensure the smooth initiation of the 240-bus miniWECC system. This method is later extended to the modeling and integration of OWFs in Section III.

\begin{figure}[htbp]
   \centering
   \subfloat[\label{fig:basic_ini}]{
        \includegraphics[scale=0.52]{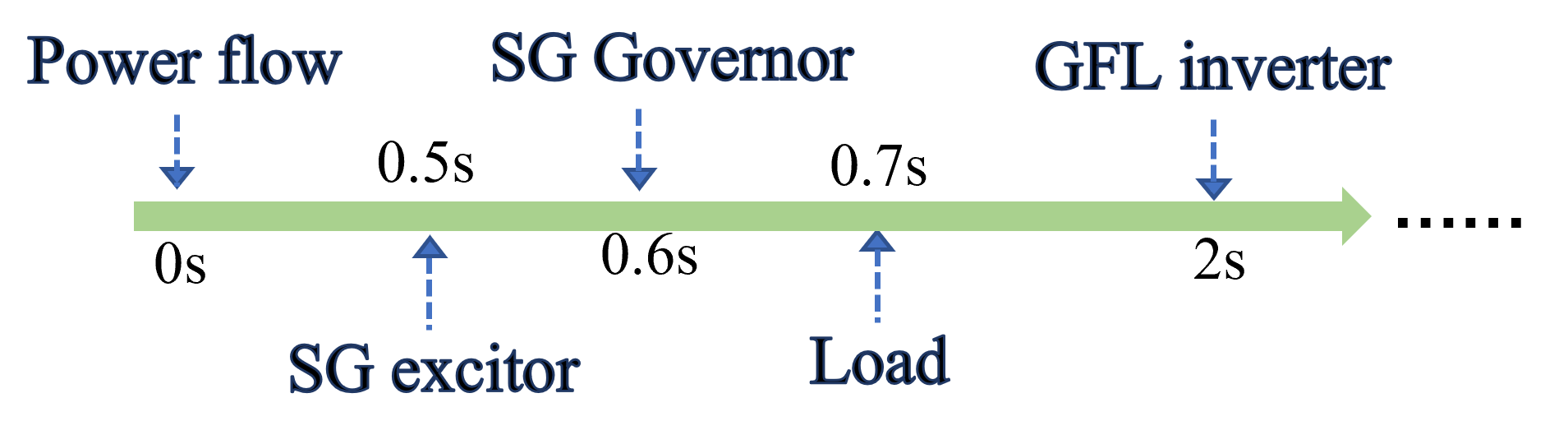}}
        \vspace{-0em} 
   \hfill
   \subfloat[\label{fig:OWF_ini}]{
         \includegraphics[scale=0.52]{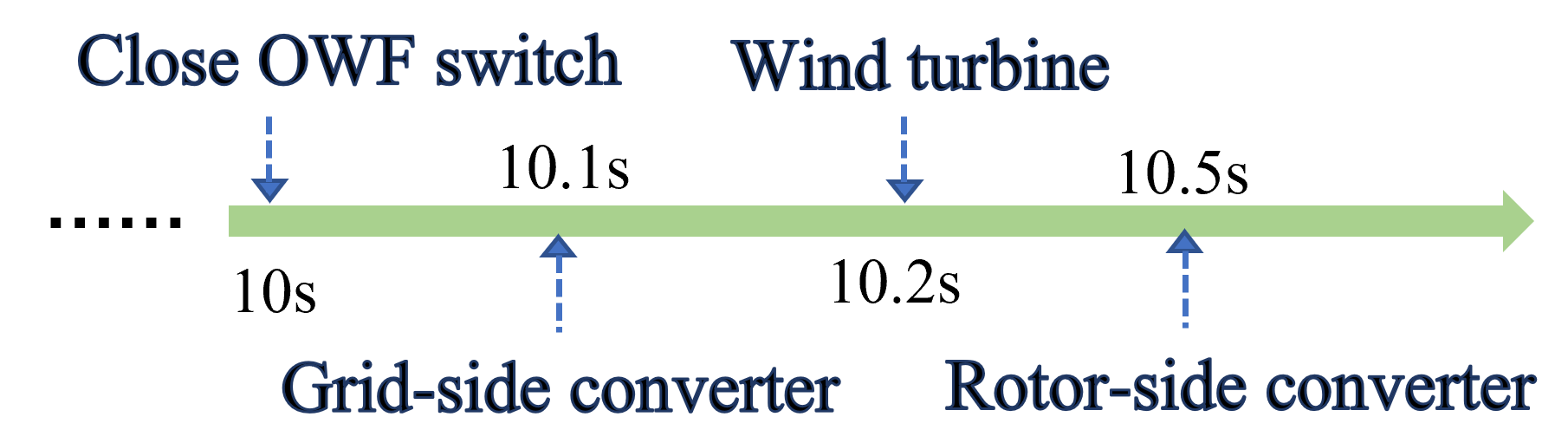}}
\caption{Sequenctial initialization for EMT simulation: (a) basic miniWECC system, (b) Off-shore wind farms.}
\label{fig:model_ini}
\end{figure}

As shown in Fig. \ref{fig:basic_ini}, the basic miniWECC system is initialized from power flow results from PSS/e. Subsequently, SG exciters come online at 0.5 s, followed by SG governors at 0.6 s. Before this stage, SGs operate in constant speed mode. At 0.7 s, constant power loads are replaced with mixed ZIP loads. The last step involves GFL IBR initialization. 

As shown in Fig. \ref{fig:GFL-IBR}, this paper parallelly connects IBRs to the ideal voltage sources converted from E-tran. They are assigned P/Q references set to 0. These reference values are then adjusted to their actual output obtained from PSS/e. Simultaneously, all breakers connecting the ideal voltage sources are opened to complete the transition from the ideal voltage sources to GFL IBRs at 2 s.

\section{Modeling, Configuring, and Integrating Offshore Wind Farms}

This section introduces the control and modeling of OWFs, followed by the configuration and initialization in PSCAD. Thereafter, the OWFs are systematically connected to the basic miniWECC system.

\subsection{Modeling and Configuration of OWFs}

\subsubsection{Modeling and control}

\begin{figure*}
    \centering
    \includegraphics[width=0.95\linewidth]{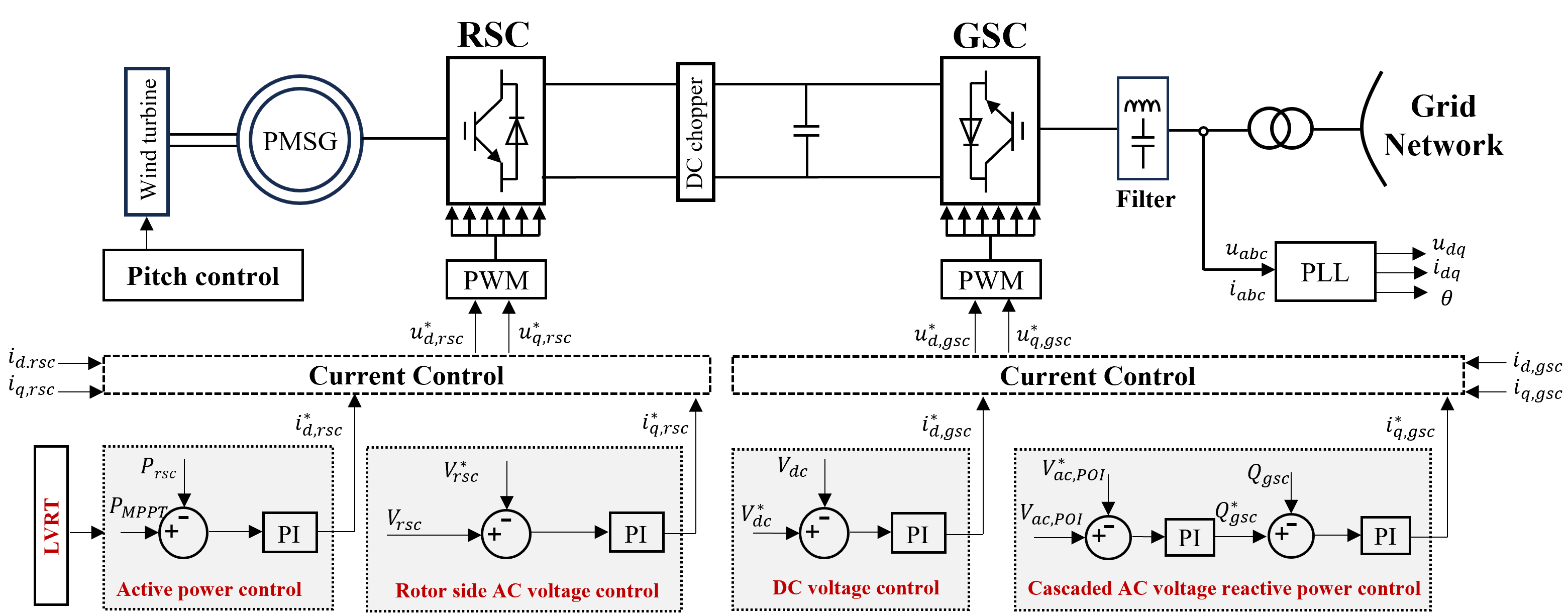}
    \vspace{0.25cm}
    \caption{Configuration of Type 4 wind turbine and its control system.}
    \label{fig:enter-label}
\end{figure*}

This paper selects the Type-4 wind turbine \cite{pscad_type4} as the basic unit of OWFs. As shown in Fig. \ref{fig:enter-label}, the wind turbine generator operates in maximum power point tracking (MPPT) modes and incorporates a pitch angle control module, a grid-side control module, a rotor-side control module, and a DC-link chopper. These control modules are briefly discussed below.

\textit{i)} The pitch angle, denoting the rotation angle of the blade around its longitudinal axis, is dynamically adjusted to ensure a safe rotational speed of the wind turbine in response to varying wind speeds. In coordination with the Yaw Control, this mechanism optimizes the turbine's performance by ensuring that the blades consistently face the wind at the optimal angle.

\textit{ii)} The grid-side control is formulated to maintain the DC voltage between the two converters and the AC voltage at the point of interconnection (POI). It generates references for an active power loop ($i_{d, gsc}^{*}$) and a reactive power loop ($Q_{gsc}^{*}$ and $i_{q, gsc}^{*}$) as illustrated in \eqref{eq:gsc}. Subsequently, the inner loops track the current reference, producing a modulation index based on \eqref{eq:inv_voltageReg} for the pulse-width modulation block.
\begin{equation}\label{eq:gsc}
\left\{\begin{array}{l}
i_{d, gsc}^{*}=\left(k_{p }+\frac{k_{i}}{s}\right)\left(V_{dc}^{*}-V_{dc}\right) \\
Q_{gsc}^{*}=\left(k_{p}+\frac{k_{i}}{s}\right)\left(V_{ac, POI}-V_{ac, POI}^{*} \right)\\
i_{q, gsc}^{*}=\left(k_{p}+\frac{k_{i}}{s}\right)\left(Q_{gsc}^{*}-Q_{gsc} \right)
\end{array}\right.
\end{equation}
\begin{equation}\label{eq:inv_voltageReg}
\left\{\begin{array}{l}
u_{d}^{*}=\left(k_{p}+\frac{k_{i}}{s}\right)\left(i_{d}^{*}-i_{d}\right)-w_n L_f i_{q}+v_{d} \\
u_{q}^{*}=\left(k_{p}+\frac{k_{i}}{s}\right)\left(i_{q}^{*}-i_{q}\right)+w_n L_f i_{d}+v_{q}
\end{array}\right.
\end{equation}

\textit{iii)} The rotor-side converter control is designed to regulate the active power and AC voltage at the terminals of the wind turbines. As depicted in \eqref{eq:rsc}, it generates $i_{d, rsc}^{*}$ based on the error between the actual rotor-side converter output and $P_{\text{MTTP}}$, and $i_{q, rsc}^{*}$ based on the error between the terminal voltage of the wind turbine and the voltage reference. Additionally, it incorporates the same current tracking loop as the grid-side converter. Note, besides MPPT, an additional active power droop lookup table is implemented in the rotor-side control to modify $P_{\text{MPPT}} $ in emergency conditions. This enhances the stability of the wind turbine generator during faults and provides low-voltage ride-through (LVRT) capability \cite{sheng2021investigation}.
\begin{equation}\label{eq:rsc}
\left\{\begin{array}{l}
i_{d, rsc}^{*}=\left(k_{p }+\frac{k_{i}}{s}\right)\left(P_{\text{MPPT}}-P_{rsc}\right) \\
i_{q, rsc}^{*}=\left(k_{p}+\frac{k_{i}}{s}\right)\left(V_{rsc}-V_{rsc}^{*} \right)
\end{array}\right.
\end{equation}

\textit{iv)} Power electronic devices have much lower overloading (over-voltage or over-current) capabilities than SGs. To protect converters during fault conditions, a DC-link chopper is configured using insulated-gate bipolar transistor (IGBT)-controlled resistance. It will be triggered once the overvoltage is detected.

\subsubsection{Sequential initilization}

Similar to the basic miniWECC system, sequential initialization is employed for OWFs. As illustrated in Fig. \ref{fig:OWF_ini}, the initialization sequence involves closing the switch, activating the grid-side converter, initiating the wind turbine, and enabling the rotor-side converter for a single OWF. Moreover, multiple OWFs can be connected to their POIs one after another, with designated time intervals, to mitigate the impact of substantial power injection on the main grid.

\subsection{Integration of OWFs into the MiniWECC System}

The technical report from PNNL \cite{travis2023} showed three strategies connecting OFWs to WECC system: 
\begin{itemize}
    \item through year 2030 HVAC radial topology;
    \item through year 2030+ HVDC radial topology;
    \item through year 2030+ MTDC backbone topology.
\end{itemize}

This paper focuses on the radial HVAC connection between OWFs and Southern Oregon onshore POIs. Specifically, the total capacity of 3.4 GW of OWFs is divided into two groups. One group of OWFs has a capacity of 0.8 GW, while the other group has a capacity of 2.6 GW. Both groups consist of wind turbines with a rated capacity of 2 MW and cut-in and cut-out wind speeds of 4 m/s and 25 m/s, respectively. These OFWs are strategically installed off the Southern Oregon coast, with the maximum power injection capped at 2.7 GW \cite{travis2023}. Fig. \ref{fig:wecc} illustrates their relative installation positions. The development of this basic use case will lay the ground for upgrading the HVAC connection to HVDC and MTDC connections in the future.

\section{Demonstration}

This section validates the configured 240-bus miniWECC system integrating OWFs in PSCAD software. Two scenarios are prepared and tested, including a flat run followed by wind speed change and a grounded fault at POI. The simulation is conducted on a computer equipped with Intel(R) Core i7-1365U running at 1.8 GHz, 16 GB of memory, and PSCAD v5.0.0 Professional. 

\subsection{Scenario 1: Flat Run and Wind Speed Variation}
\begin{figure}[htbp]
   \centering
   \subfloat[\label{fig:w_speed}]{
        \includegraphics[scale=0.68]{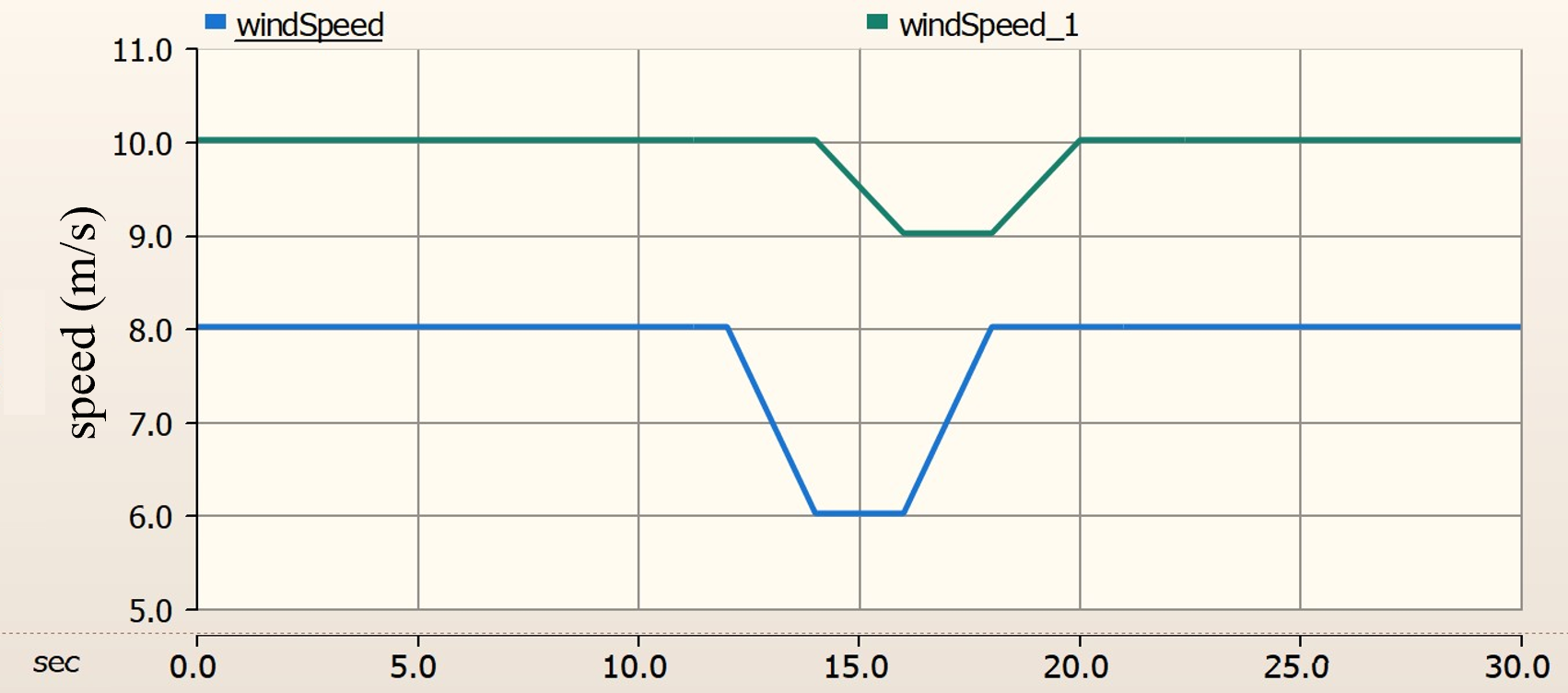}}
        \vspace{-0em} 
   \hfill
   \subfloat[\label{fig:p_s1}]{
         \includegraphics[scale=0.68]{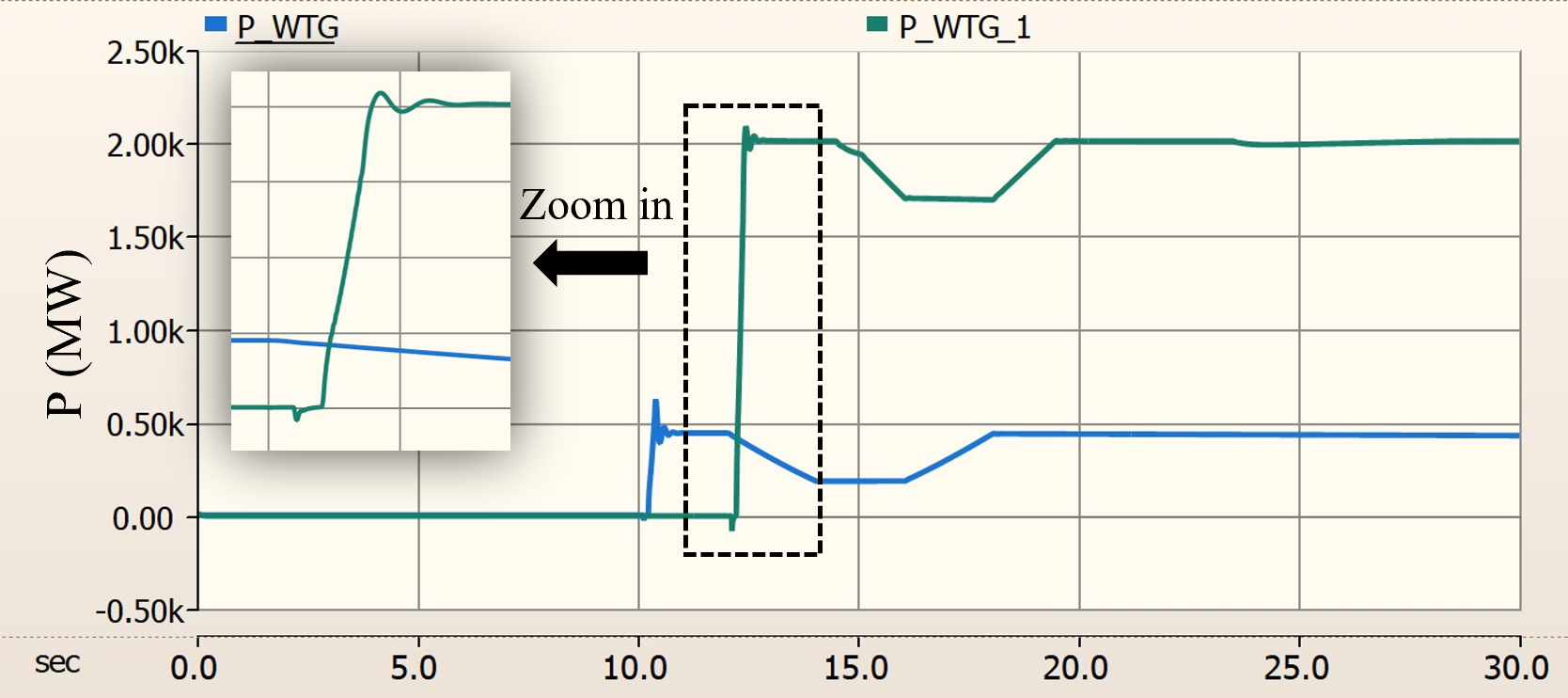}}
\caption{Dynamic responses of OWFs in Scenario 2: (a) wind speed variation; (b) active power injection.}
\label{fig:dynamic_s1}
\end{figure}

In Scenario 1, the configured system is initialized in sequence according to the order shown in Fig. \ref{fig:model_ini}. Then, two distinct wind speed curves in Fig. \ref{fig:w_speed} are applied to the two OWFs. 

Fig. \ref{fig:dynamic_s1} shows the stable active power response. The first OWF connects to the main grid at 10 s, followed by the second one at 12 s. Then, their output adjusts in response to wind speed variations until t=20 s. The results demonstrate that the OWFs can seamlessly connect to the miniWECC system, operating in MPPT mode and adapting to changes in wind speed.

\subsection{Scenario 2: Grounded Faults}

To better observe the dynamics of OWFs under grounded faults in Scenario 2, the wind speed is assumed constant at 10m/s. 
A three-phase grounded fault happens at the POI of the 2.6 GW OWF at 15 s. The fault is cleared at 0.15 s. 

\begin{figure}[htbp]
   \centering
   \subfloat[\label{fig:V_fault}]{
         \includegraphics[scale=0.63]{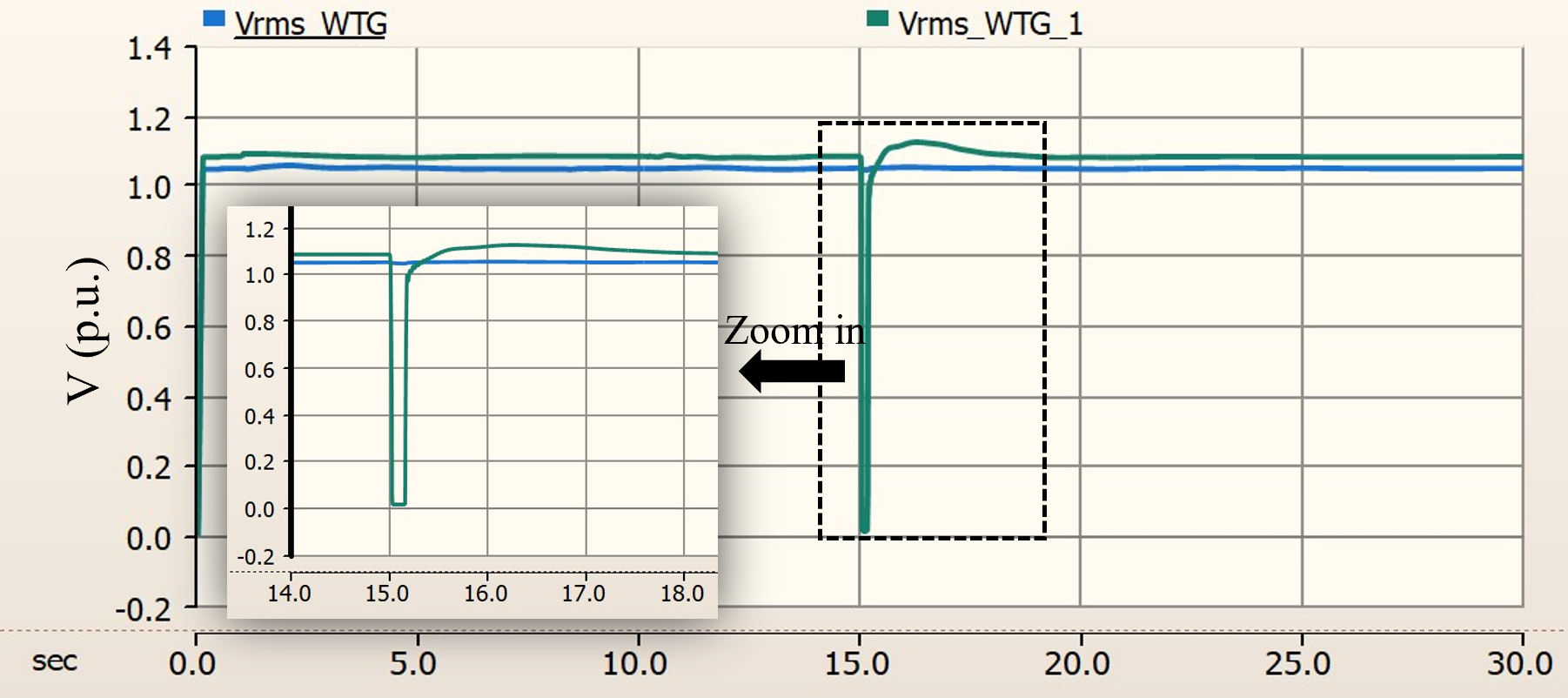}}
   \hfill
   \subfloat[\label{fig:P_fault}]{
        \includegraphics[scale=0.64]{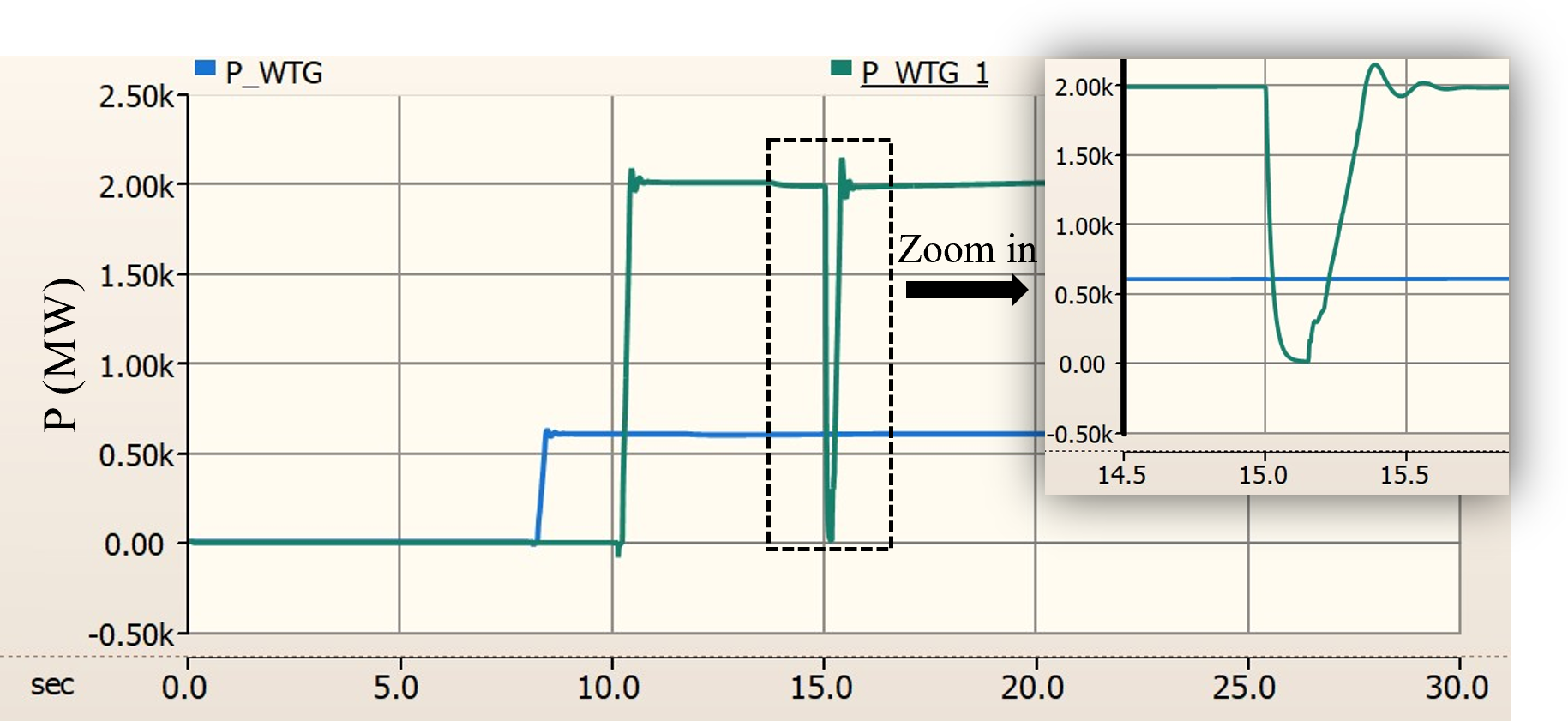}}
        \vspace{-0em} 
\caption{Dynamic responses of OWFs in Scenario 2: (a) voltage of POI; (b) active power injection.}
\label{fig:dynamic_s2}
\end{figure}

Fig. \ref{fig:dynamic_s2} shows the dynamic response of the OWF. At 15 s, the POI AC voltage in Fig. \ref{fig:V_fault} drops to 0 due to the grounded fault. Because the OWF is equipped with a DC-link chopper, its DC capacitor has no obvious overvoltage and thus does not trip the OWF during the fault. The AC voltage recovers instantly after the fault is cleared and comes back to its nominal value around t=20 s. Accordingly, the active power injection of OWF drops significantly during the fault, and it recovers to the value of 2 GW around 15.5 s. The dynamic results demonstrate that the configured OWF is robust to ground faults when connected to the miniWECC system.

\section{Conclusion}
This paper illustrates the incorporation of OWFs into the 240-bus miniWECC system and its configuration in PSCAD software. Simulation results show that OWFs are adaptive to wind speed variation and robust under grounded faults. The system lays the ground for modeling the HVAC and MTDC in the WECC system. It also serves as a use case for validating the fast dynamic performance of future WECC systems with high penetration of wind energy.

\section*{Acknowledgment}
The authors want to thank Kaustav Chatterjee for his comments and suggestions to improve this work. This work is funded by the Laboratory Directed Research and Development (LDRD) at PNNL as part of the E-COMP initiative. PNNL is operated by Battelle for the DOE under Contract DOE-AC05-76RL01830.

\bibliographystyle{IEEEtran}
\bibliography{1bib} 

\end{document}